\documentstyle[psfig]{elsart}

\begin{document}

\begin{frontmatter}

\title{Electron-Phonon Interaction in $NbB_{2}$ : A 
Comparison with  $MgB_{2}$}

\author{Prabhakar P. Singh}

\address{Department of Physics, Indian Institute of Technology, Powai, Mumbai- 400076,
India}

\begin{abstract}
We present a comparison of electron-phonon interaction in \( NbB_{2} \) and
\( MgB_{2} \), calculated using full-potential, density-functional-based methods
in \( P6/mmm \) crystal structure. Our results, described in terms of (i) electronic
structure, (ii) phonon density of states \( F(\omega ) \), (iii) Eliashberg
function \( \alpha ^{2}F(\omega ) \), and (iv) the solutions of the isotropic
Eliashberg gap equation, clearly show significant differences in the electron-phonon
interaction in \( NbB_{2} \) and \( MgB_{2} \). We find that the average electron-phonon
coupling constant \( \lambda  \) is equal to \( 0.59 \) for \( MgB_{2} \)
and \( 0.43 \) for \( NbB_{2} \), leading to superconducting transition temperature
\( T_{c} \) of around \( 22\, K \) for \( MgB_{2} \) and \( 3\, K \) for
\( NbB_{2}. \)
\end{abstract}
%%\pacs{PACS numbers: 74.25.Jb, 74.70.Ad}
\end{frontmatter}

The lack of success in finding superconductivity in other diborides with superconducting
transition temperature, \( T_{c} \), close to that of \( MgB_{2} \) \cite{akimitsu}
underscores the complex nature of interaction responsible for superconductivity
in \( MgB_{2}. \) In \( MgB_{2} \) the complexity is further compounded by
the presence of multi-faceted Fermi surface \cite{kortus,joon1} and a highly
anisotropic electron-phonon coupling, \( \lambda (\mathbf{k},\mathbf{k}'), \)
 over the Fermi surface \cite{kong,joon2}. The dependence of superconducting
properties on such details has ensured that we do not know, as yet, the exact
nature of interaction leading to superconductivity in \( MgB_{2}. \) 

Within Eliashberg-Migdal theory \cite{allen1,allen2} of superconductivity,
a reliable description of the superconducting state requires an accurate knowledge
of \( \lambda (\mathbf{k},\mathbf{k}') \) and the renormalized electron-electron
interaction, \( \mu ^{*} \), which are used as input to the fully anisotropic
gap equation. The present computational capability allows us to evaluate \( \lambda (\mathbf{k},\mathbf{k}') \)
accurately using density-functional-based methods but, unfortunately, \( \mu ^{*} \)
cannot be evaluated. However, it is reasonable to assume that \( \mu ^{*} \)
varies between \( 0.1 \) to \( 0.2 \) \cite{kong,joon2}. Thus, the electron-phonon
coupling \( \lambda (\mathbf{k},\mathbf{k}') \), which is a normal state function,
must contain signatures of superconducting state. 

In an attempt to identify some of the unique features of electron-phonon interaction
in \( MgB_{2} \) \emph{vis-a-vis} other diborides we have studied (i) the electronic
structure, (ii) the phonon density of states (DOS), (iii) the Eliashbrg 
function, and (iv) the solutions of the isotropic Eliashberg gap equation
for \( NbB_{2} \) and \( MgB_{2} \) in \( P6/mmm \) crystal structure. 

The choice of \( NbB_{2} \) has been motivated by the recent reports \cite{yamamoto,akimitsu2}
of superconductivity, with \( T_{c} \) going up to \( 9.2\, K \), under pressure
in hole-doped \( Nb_{x}B_{2} \). Earlier experiments have shown superconductivity
in stoichiometric \( NbB_{2} \) \cite{akimitsu2,layrovska} as well as Boron-enriched
\( NbB_{2} \) \cite{cooper} samples. The reported \( T_{c} \) for stoichiometric
\( NbB_{2} \) varies from \( 0.62\, K \) \cite{layrovska} to \( 5.2\, K \)
\cite{akimitsu2}, while for Boron-enriched \( NbB_{2.5} \) the \( T_{c} \)
is found to be \( 6.4\, K \) \cite{cooper}. We also note that, recently, Kaczorowski
\emph{et al.} \cite{kaczorowski} did not find any superconductivity in \( NbB_{2} \) down to \( 2\, K. \)

We have calculated the electronic structure of \( NbB_{2} \) and \( MgB_{2} \)
in \( P6/mmm \) crystal structure with optimized lattice constants \( a \)
and \( c \), as given in Table I. The lattice constants \( a \) and \( c \)
were optimized using the ABINIT program \cite{abinit} based on pseudopotentials
and plane waves. For studying the electron-phonon interaction we used the full-potential
linear response program of Savrasov \cite{savrasov1,savrasov2}, and   calculated
the dynamical matrices and the Hopfield parameter. These  were then used to calculate
the phonon DOS, \( F(\omega ) \), the electron-phonon coupling \( \lambda (\mathbf{k},\mathbf{k}') \),
and the Eliashberg function, \( \alpha ^{2}F(\omega ) \), for \( NbB_{2} \)
and \( MgB_{2} \). Subsequently, we have numerically solved the isotropic Eliashberg
gap equation \cite{allen1,allen2,private} for a range of \( \mu ^{*} \) to
obtain the corresponding \( T_{c}. \) 

Based on our calculations, described below, we find significant differences
in the phonon DOS and the Eliashberg functions of \( NbB_{2} \) and \( MgB_{2} \).
In particular, we find that the average electron-phonon coupling constant is
equal to \( 0.59 \) for \( MgB_{2} \) and \( 0.43 \) for \( NbB_{2} \),
 leading to superconducting transition temperatures of around \( 22\, K \) for
\( MgB_{2} \) and \( 3\, K \) for \( NbB_{2}. \)

Before describing our results in detail, we provide some of the computational
details of our calculation. The structural relaxation was carried out by the molecular dynamics
program ABINIT \cite{abinit} with Broyden-Fletcher-Goldfarb-Shanno minimization
technique using Troullier-Martins pseudopotential \cite{troullier} for \( MgB_{2} \)
and Hartwigsen-Goedecker-Hutter pseudopotential \cite{hgh} for \( NbB_{2} \),
512 Monkhorst-Pack \cite{monkhorst} \( k \)-points and Teter parameterization
for exchange-correlation. The kinetic energy cutoff for the plane waves was
\( 110\, Ry \) for \( MgB_{2} \) and \( 140\, Ry \) for \( NbB_{2}. \) The
charge self-consistent full-potential LMTO \cite{savrasov1} calculations were
carried out with the generalized gradient approximation for exchange-correlation
of Perdew \emph{et al} \cite{perdew} and 484 \( k \)-points in the irreducible
wedge of the Brillouin zone. For \( MgB_{2} \), the basis set used consisted
of $3\kappa$ panels and \( s, \) \( p, \) \( d \) and \( f \) orbitals at the \( Mg \) site
and \( s, \) \( p \) and \( d \) orbitals at the \( B \) site. In the case
of \( NbB_{2} \), we included $2\kappa$ panels and \( s \), \( p \) and \( d \) orbitals at the
\( Nb \) site. In all cases the potential and the wave function were expanded
up to \( l_{max}=6 \). The muffin-tin radii for \( Mg \), \( B \), and \( Nb \)
were taken to be \( 2.4, \) \( 1.66, \) and \( 2.3 \) atomic units, respectively. 

The calculation of dynamical matrices and the Hopfield parameters for \( MgB_{2} \)
were carried out using a \( 6\times 6\times 6 \) grid while for \( NbB_{2} \) we used
a \( 4\times 4\times 4 \) grid resulting in \( 28 \) and \( 12 \) irreducible \( \mathbf{q} \)-points,
respectively. For Brillouin zone integrations in \( MgB_{2} \) we used a \( 6\times 6\times 6 \)
grid while for \( NbB_{2} \) we used \( 8\times 8\times 8 \) grid of \( \mathbf{k} \)-points.
The Fermi surface was sampled more accurately with a \( 24\times 24\times 24 \) grid of
\( \mathbf{k} \)-points using the double grid technique as outlined in Ref.
\cite{savrasov2}. 

Here, we like to point out the reasons for carrying out the linear response calculation
for \( MgB_{2} \) in spite of earlier calculations by Kong \emph{et al}. \cite{kong}
and Choi \emph{et al.} \cite{joon1,joon2}. \emph{}The linear response calculation
by Kong \emph{et al}. is similar to the present approach, while Choi \emph{et
al}. used pseudopotentials and frozen phonon method to evaluate the electron-phonon
coupling \( \lambda (\mathbf{k},\mathbf{k}') \). The present approach differs
from the work of Kong \emph{et al} in the selection of \textbf{q}-points and
the Brillouin zone integrations. As a result Kong \emph{et al}. find the average
electron-phonon coupling constant \( \lambda =0.87\pm 0.05 \), which is much
higher than the value of \( \lambda =0.61 \) as reported by Choi \emph{et al}.
\cite{joon1,joon2}, as well as the experimental values of  \( 0.58 \)
\cite{wang} and \( 0.62 \) \cite{bouquet} as deduced from specific heat measurements.
Our calculated value of \( \lambda  \) is equal to \( 0.59, \) in close agreement
with the work of Choi \emph{et al}..
\begin{table}

\caption{The calculated lattice constants \protect\( a\protect \) and \protect\( c.\protect \)
The experimental lattice constants for \protect\( MgB_{2}\protect \) \cite{lipp}
and \protect\( NbB_{2}\protect \) are shown in the parentheses.}
{\centering \begin{tabular}{|c|c|c|}
\hline 
&
\( a \) (a.u.)&
\( c \) (a.u.)\\
\hline 
\hline 
\( MgB_{2} \)&
5.76 (5.834)&
6.59 (6.657)\\
\hline 
\( NbB_{2} \)&
5.81 (5.837)&
6.10 (6.245)\\
\hline 
\end{tabular}\par}\end{table}

\begin{table}

\caption{The site- and \protect\( l\protect\)-resolved electronic densities of states,
in \protect\( st/(Ry-atom)\protect \), at the Fermi energy in \protect\( MgB_{2}\protect \)
and \protect\( NbB_{2}\protect \) calculated at the optimized lattice constants
using the full-potential LMTO method. }
{\centering \begin{tabular}{|c|c|c|c|c|c|}
\hline 
alloy&
element&
\( s \)&
\( p \)&
\( d \)&
\( f \)\\
\hline 
\hline 
\( MgB_{2} \)&
\( Mg \)&
0.47&
0.72&
0.94&
0.08\\
\hline 
&
\( B \)&
0.06&
3.36&
0.15&
-\\
\hline 
\( NbB_{2} \)&
\( Nb \)&
0.02&
.09&
9.54&
- \\
\hline 
&
\( B \)&
0.07&
1.34&
0.19&
-\\
\hline 
\end{tabular}\par}\end{table}

A comparison of site- and \( l \)-resolved electronic density of states of
\( NbB_{2} \) \cite{pps1} and \( MgB_{2} \) \cite{pps2} at the Fermi energy
is given in Table II. A further decomposition of the densities of states ($st/Ry$)  in
terms of cubic harmonics reveals the dominance of \( B \) \( p \) electrons
at the Fermi energy in \( MgB_{2} \) \( (p_{x(y)}=0.9,\,  p_{z}=1.55) \)
than in \( NbB_{2} \) \( (p_{x(y)}=0.43,\, p_{z}=0.48) \).
In addition, in \( NbB_{2} \), the \( Nb \) \( d \)-electrons \( (d_{xy(x^{2}-y^{2})}=1.83,\, d_{yz(zx)}=1.66,\, d_{3z^{2}-1}=2.56) \)
are present in substantial amount, indicating a more active role for \( Nb \)
in determining the possible superconducting properties of these materials than
played by \( Mg \) in \( MgB_{2}. \) 

\begin{figure}
{\par\centering 
\psfig{file=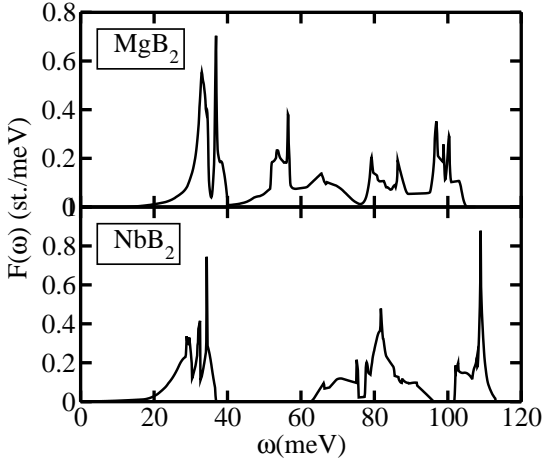,height=7.4cm,angle=-90} 
\par}

\caption{The phonon density of states \protect\( F(\omega )\protect \) of \protect\( MgB_{2}\protect \)
and \protect\( NbB_{2}\protect \) calculated using the full-potential linear
response method as described in the text.}
\end{figure}

In Fig. 1 we show the phonon DOS \( F(\omega ) \) of \( NbB_{2} \) and \( MgB_{2} \)
calculated using the full-potential linear response program as described earlier.
For \( MgB_{2} \) we can clearly identify four significant peaks in the phonon
DOS at \( 33, \) \( 53, \) \( 79 \) and \( 96 \) \( meV \), respectively.
The peak at \( 33 \) \( meV \) is related to the van Hove singularity in the
acoustical mode \cite{joon2}, and it involves the motion of \( Mg \) atom
and \( B \) atoms separately. However, the region around the peak at \( 53 \)
\( meV \) results from the motion of both \( Mg \) and \( B \) atoms. The
phonon DOS around \( 79 \) and \( 96 \) \( meV \) peaks are due to the coupled
motion of \( B-B \) atoms in the \( x-y \) plane. In particular, the peak
at \( 79 \) \( meV \) corresponds to the in-plane \( B-B \) bond stretching
mode, and in Ref. \cite{joon2} it is located at  \( 77\, meV \). Similarly,
for \( NbB_{2} \), the peak in \( F(\omega ) \) at \( 32 \) \( meV \) is
dominated by the motion of \( Nb \) atom, while the region around \( 65-70 \)
\( meV \) results from the coupled motion of \( Nb \) and the two \( B \)
atoms. Not surprisingly, \( 81 \) \( meV \) peak in the phonon DOS of \( NbB_{2} \)
corresponds to the in-plane \( B-B \) motion. In contrast with the phonon DOS
in \( MgB_{2}, \) the phonon DOS in \( NbB_{2} \) around \( 106 \) \( meV \)
results from the displacements of both \( Nb \) and the two \( B \) atoms. 

\begin{figure}
{\par\centering 
\psfig{file=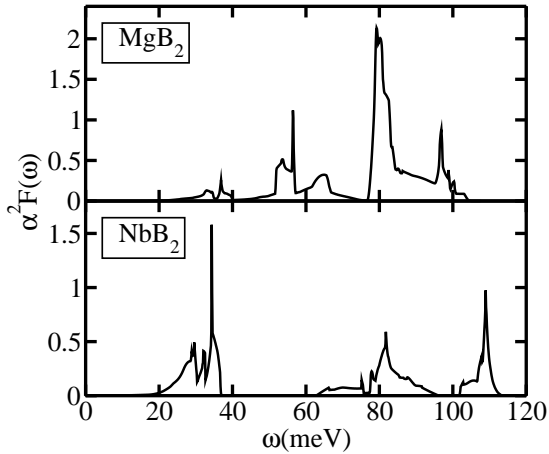,height=7.4cm,angle=-90} 
\par}

\caption{The Eliashberg function \protect\( \alpha ^{2}F(\omega )\protect \) of \protect\( MgB_{2}\protect \)
and \protect\( NbB_{2}\protect \) calculated using the full-potential linear
response method as described in the text.}
\end{figure}

To see the strengths with which the different modes of the ionic motion couple
to the electrons, and thus are capable of influencing the superconducting properties
the most, we show in Fig. 2 the Eliashberg function \( \alpha ^{2}F(\omega ) \)
of \( NbB_{2} \) and \( MgB_{2} \) calculated as described earlier. The most
striking feature of Fig. 2 is the overall strength of the electron-phonon coupling
in \( MgB_{2} \) as compared to \( NbB_{2} \). We find that the average electron-phonon
coupling constant \( \lambda  \) is equal to \( 0.59 \) for \( MgB_{2} \)
and \( 0.43 \) for \( NbB_{2} \), which clearly shows that \( MgB_{2} \)
is more likely to show superconductivity with a higher \( T_{c} \) than \( NbB_{2}. \)

Further analysis of the Eliashberg function, as shown in Fig. 2, reveals the
importance of the in-plane \( B-B \) bond-stretching optical phonon mode in
\( MgB_{2} \), which gives rise to the dominant peak at \( 79\, meV \). The
other peaks in the phonon DOS of \( MgB_{2} \), such as the peaks at \( 33 \)
and \( 53 \) \( meV \), couple weakly with the electrons at the Fermi energy.
Thus the motion of \( Mg \) atom plays a relatively insignificant role in determining
the superconducting properties of \( MgB_{2}. \) In contrast, in the case of
\( NbB_{2} \) the phonon modes with peaks at \( 32 \), \( 81 \) and \( 106 \)
\( meV \) couple to the electrons with almost equal strength, albeit much smaller
than in \( MgB_{2}, \) as can be seen from Fig. 2. We could have expected this
because of the significant presence of the \( Nb \) \( d \) electrons at the
Fermi energy. In Table III we have listed the Hopfield parameter \( \eta , \)
the electron-phonon coupling constant \( \lambda  \), and the various averages
of the phonon frequencies for \( NbB_{2} \) and \( MgB_{2} \). The values
listed in Table III for \( MgB_{2} \) are in good agreement with the corresponding
results of Choi \emph{et al.} \cite{joon2}.

\begin{table}

\caption{The calculated Hopfield parameter \protect\( \eta \protect \), the average
electron-phonon coupling constant \protect\( \lambda \protect \), the root
mean square \protect\( <\omega ^{2}>^{1/2}\protect \) and the logarithmically
averaged \protect\( \omega _{ln}\protect \) phonon frequencies for \protect\( MgB_{2}\protect \)
and \protect\( NbB_{2}\protect \).}
{\centering \begin{tabular}{|c|c|c|c|c|}
\hline 
 alloy&
\( \eta \, (mRy/a.u.^{2}) \)&
\( \lambda  \)&
\( <\omega ^{2}>^{1/2}\, (K) \) &
\( \omega _{ln}\, (K) \)\\
\hline 
\hline 
\( MgB_{2} \)&
167&
0.59&
835&
768\\
\hline 
\( NbB_{2} \)&
203&
0.43&
669&
494\\
\hline 
\end{tabular}\par}\end{table}

To examine the superconducting transition temperature, \emph{if} any, \emph{}of
\( NbB_{2} \) and \( MgB_{2} \) we have used the calculated Eliashberg function
\( \alpha ^{2}F(\omega ) \) to solve numerically the isotropic gap equation
\cite{allen2,private}, and the results are shown in Fig. 3 for a range of values
of \( \mu ^{*}. \) From Fig. 3 we find that for \( \mu ^{*}=0.1 \) the \( T_{c} \)
for \( MgB_{2} \) is equal to \( \sim 23\, K \), while for \( NbB_{2} \)
it is equal to \( \sim 4\, K \). Thus, our calculation shows that \( NbB_{2} \)
is superconducting with a possible \( T_{c} \) of around \( 3\, K. \) It is
worthwhile to point out that to obtain a \( T_{c} \) close to that of \( 39\, K \)
for \( MgB_{2} \), as found experimentally \cite{akimitsu}, one has to solve
the \emph{anisotropic} gap equation \cite{joon2}. The need to solve the anisotropic
gap equation for \( MgB_{2} \) rather than the isotropic gap equation arises
due to the highly anisotropic electron-phonon coupling \( \lambda (\mathbf{k},\mathbf{k}') \)
\cite{joon1,kong,joon2} over the Fermi surface. In the case of \( NbB_{2} \)
the electron-phonon coupling is \emph{neither as strong nor as anisotropic},
and thus the results obtained with the isotropic gap equation are reliable. 

As indicated earlier, the experiments show superconductivity in hole-doped \( Nb_{x}B_{2} \)
under pressure and Boron-enriched \( NbB_{2}. \) Hole-doping and Boron enriching
both lead to a relative increase in the Boron population at the Fermi energy
and, probably, enhances the peak around \( 81\, meV \) leading to an increase
in \( T_{c}. \) Of course a more quantitative investigation is needed to pinpoint
the exact nature of changes in \( NbB_{2} \) which lead to superconductivity.

In conclusion, we have studied the electron-phonon interaction in \( NbB_{2} \)
and \( MgB_{2} \), using full-potential, density-functional-based methods in
\( P6/mmm \) crystal structure. We have described our results in terms of (i)
electronic structure, (ii) phonon density of states, (iii) Eliashberg function,
and (iv) the solutions of the isotropic Eliashberg gap equation, which clearly
show significant differences in the electron-phonon interaction in \( NbB_{2} \)
and \( MgB_{2} \). We find that the average electron-phonon coupling constant
is equal to \( 0.59 \) for \( MgB_{2} \) and \( 0.43 \) for \( NbB_{2} \),
leading to superconducting transition temperature of around \( 22\, K \) for
\( MgB_{2} \) and \( 3\, K \) for \( NbB_{2}. \)

\begin{figure}
{\par\centering 
\psfig{file=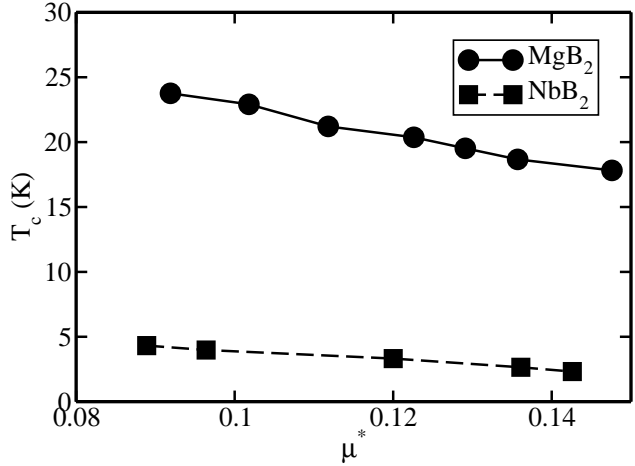,height=7.4cm,angle=-90} 
\par}

\caption{The superconducting transition temperature \protect\( T_{c}\protect \) as
a function of \protect\( \mu ^{*}\protect \) for \protect\( MgB_{2}\protect \)
and \protect\( NbB_{2}\protect \) as obtained from the isotropic Eliashberg
gap equation.}
\end{figure}

\end{document}